\newcommand{\diracslash}[1]{#1\llap{/\kern2pt}}

\newcommand{\be}{\begin{equation}}
\newcommand{\ee}{\end{equation}}
\newcommand{\bea}{\begin{eqnarray}}
\newcommand{\eea}{\end{eqnarray}}
\newcommand{\ba}[1]{\begin{array}{#1}}
\newcommand{\ea}{\end{array}}

\documentclass[prd,aps,floats,nofootinbib,tightenlines]{revtex4}
\usepackage{epsfig,graphicx,pstricks}
\usepackage{psfrag}
\usepackage{color}
\usepackage{amsmath}
\usepackage{amsfonts}
\usepackage{amssymb}
\usepackage{color, soul}
\usepackage{textcomp}
\usepackage{placeins}
\usepackage{multirow}
\usepackage{natbib}
\usepackage{subfigure}
\usepackage{hyperref}
\begin{document}

\title{Study of Gas Electron Multiplier Detector Using ANSYS and GARFIELD$^{++}$}
\author{Md Kaosor Ali Mondal}
\email{mohammad.mondal@cern.ch}
\author{Poojan Angiras}
\email{poojan.angiras@cern.ch}
\author{Sachin Rana}
\email{sachin.rana@cern.ch}
\author{A.~Sarkar}
\email{amal.sarkar@cern.ch}

\affiliation{School of Physical Science, Indian Institute of Technology Mandi, Kamand, Mandi - 175005, India}

\begin{abstract}
Micro-Pattern Gas Detectors (MPGDs) represent a category of gaseous ionization detectors that utilize microelectronics. They feature a remarkably small distance between the high potential difference anode and cathode electrodes and are typically filled with gases. When a high-energy particle interacts with the gas medium, it generates ions and electrons, which are subsequently accelerated in opposite directions due to the applied electric field. Deflected electrons trigger further ionization to create electron-ion pairs through an avalanche process. These particles can be detected with very high precision at the readout. The Gas Electron Multiplier (GEM) is one type of MPGD constructed with a polyimide film sandwiched between two conductors under a high voltage difference. Microscopic holes in the foil facilitate electron avalanche. However, the current geometry of the GEM  detector used in various experiments is sub-optimal for the gain and performance. In this study, we have modified the geometry of the GEM detector to enhance the gain, reduce ions backflow, and enhance the performance of the detector. We are proposing a new geometry of the GEM detector foil for higher gain, better performance, and durability. For this study, the geometry has been constructed in ANSYS, and further studies have been performed using Garfield$^{++}$. 
\end{abstract}

\maketitle

\section{Introduction}
In 1895, when Wilhelm Rontgen discovered X-rays \cite{panchbhai2015wilhelm}, the development of various types of detectors began to analyze the properties of radiation and its interaction with matter. In the past, the detectors were only capable of counting the particles that were passing through them. But now the detectors can count the particles with their energy and momentum, measure their velocity, and track their path. In a gaseous detector, the device is filled with counting gas. Two electrodes are used, which are separated by the gas medium. The gas became ionized to absorb energy from the radiative particle or high-energy particle. A high potential difference is needed for the detector to become functional. In more recent years (1968), the invention of the Multi Wire Proportional Chamber (MWPC) \cite{charpak1970some} at CERN brought an evolution to the gas detector. This detector provides a breakthrough in particle detection by tracking particles \cite{lin1985novel} and reading their energy. The inventor of the MWPC, Georges Charpak, was awarded the Nobel Prize in 1992. Despite being successfully used in particle physics research and other domains, MWPCs are fundamentally limited in many ways like limited rate capability and low time resolution. A new type of gas detector is needed to improve the rate capability and position resolution of the radiative particle. 

\par
The Gas Electron Multiplier (GEM) \cite{bouclier1997gas} is one type of Micropattern Gaseous Detector (MPGD) which was developed by F. Sauli at the CERN laboratory in 1997. GEM is a gaseous ionization detector used in nuclear and particle physics to detect charged particles like X-rays, alpha, and beta, and it can also measure their energy, momentum, and spatial resolution. Two 5 \(\mu\)m copper foils are separated by a 50 \(\mu\)m kapton foil. Using the photolithography \cite{cheng2004combined} technique, several holes are created in the foil. The holes are created in the honeycomb structure to reduce the amount of unused space. Due to this spatial structure, a very high electric field is created inside the holes, causing electron avalanches. The detector is filled with two types of gases: argon and carbon dioxide. Argon is used as an ionizing gas because its ionization potential is lower (15.6 eV) than that of other inert gases. Carbon dioxide is used as a quenching gas. After the recombination of ions and electrons, a photon can be emitted, which leads to an unwanted (fake) signal. A quencher is used to absorb these types of photons. The upper side of the detector is connected to the negative terminal of the source and acts as a cathode. The lower side of the detector is connected to the positive terminal of the source, and it acts as an anode. The copper layers are kept at the potential difference of 300V. \\

In this study, we have modeled GEM detector with different foil geometries to enhance the gain and reduce the ion backflow of the detector to enhance the performance of the detector. The modeling of GEM is done with ANSYS Maxwell for calculating the electric field in every region of the detector. Then with the help of ANSYS Mechanical APDL, unit geometry of GEM is created which is used for further detailed study using Garfield$^{++}$. The following section provides a brief overview of the working principle of the GEM. Section III goes through the operational parameters of the detector. Section IV details the step-by-step construction of the GEM foil using ANSYS, while Section V outlines the technical tools used in this study. In Section VI, the performance of the detector is analyzed for various GEM foil geometries. Finally, all the results have been summarized and concluded.

\begin{figure*}
  \centering
   {{\includegraphics[width=5.2cm]{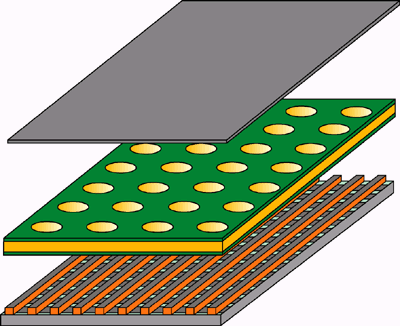}} }
    \caption{(color online) \cite{imran2019sissa} Model of Gas Electron Multiplier with kapton sandwiched between two thin copper layers with high density holes kept in the region with high potential difference.}
    \label{fig:1}
\end{figure*}

\section{Working Principle of Gas Electron Multiplier}

The Gas Electron Multiplier (GEM) detector operates through distinct regions with varying electric field strengths that control the movement of ionized particles. This section focuses on three key areas of the detector: the drift region, where initial ionization occurs as high-energy particles pass through; the avalanche region, where a strong electric field inside the GEM holes accelerates electrons, causing further ionization; and the induction region, where electrons are collected by the anode, enabling precise particle detection. Understanding these regions is crucial for optimizing the performance and efficiency of GEM detectors.

\subsection{Drift region}
The region between the cathode and the upper side of the copper foil is called the drift region. When a high-energy particle passes through the drift region, it collides with the molecules and atoms in the gas medium, and in every collision, it transfers some energy to the molecules of the gas medium. This energy is absorbed by electrons from the outermost shells of the molecules. If the energy is sufficient, the electrons absorb the energy and are free from the molecules. Thus, positive ions and negative electrons are created. This type of collision occurs many times along the trajectory of the high-energy particle, and the primary ions and electrons are created. Due to the electric field electrons move toward the gem hole. The drift region length is 3 mm, which is needed to ionize the gas molecules. Due to low electric field strength(60–70 V/mm), no electron avalanche takes place in this region.


\subsection{Avalanche}
The electric field strength inside the hole is very high (3–4 kV/mm) therefore electron momentum increases and collides with gas molecules with higher energy resulting in secondary, tertiary, and further ionization which leads to electron avalanche. In Figure 3, the avalanche process which takes place in holes of the GEM foil is shown. The studies have been performed using Garfield$^{++}$.


\begin{figure*}
  \centering
   {{\includegraphics[width=8.2cm]{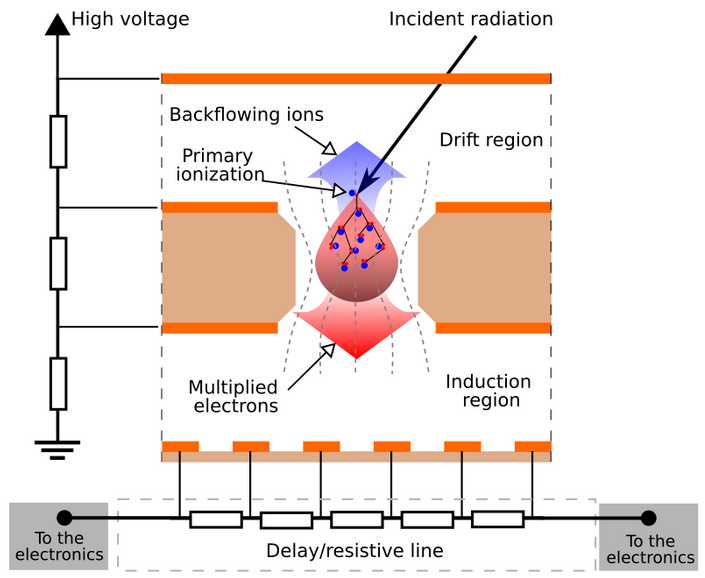} }}
    \caption{(color online) Schematic diagram of GEM, showing the primary and secondary ionization in the hole region. Multiplied electrons move towards the induction region which is then sent to the electronics and backflowing ions move towards the Drift region which contributes to the back current.}
    \label{fig:1}
\end{figure*}

\subsection{Induction region}
The region between the lower side of the copper foil and the anode is called the induction region. In this region, produced electrons moved towards the anode. The standard operating electric field strength inside the region is 100–300 V/mm. The length of this region is 2 mm, which reduces the diffusion \cite{bachmann2002discharge} of the electrons and increases the spatial resolution. 
After creating a large number of electrons during the avalanche process inside the hole, the electrons move towards the anode through this region and the ions move backward in the influence of the electric field. Because the electric field strength is small inside this region, further avalanches are minimal.
The majority of ions are collected by the upper copper layer of the GEM foil and electrons are collected by the anode which acts as a readout plate. When electrons and ions complete their cycle, they generate a current pulse \cite{malhotra2018various}. By analyzing this current pulse, the properties of the incoming high-energy particle can be studied.

\section{Operational parameters of the GEM}
The efficiency of the foil and detector in a GEM detector is influenced by two key factors: gain and ion backflow. Gain refers to the electron multiplication that occurs inside the holes of the GEM foil, resulting in a measurable current pulse. Ion backflow occurs when ions move toward the cathode after electron avalanches, reducing the overall current and causing disturbances in the detector. Both of these parameters play a crucial role in optimizing the detector's performance for high-energy particle detection.
\subsection{Gain}
In the GEM detector, electron multiplication occurs inside the holes. The number of electrons produced after avalanches for a single electron entering into a hole has been considered as the gain. 
Gain increases when the maximum number of electrons reaches the anode. A large current pulse is obtained from the particular high energy particle. Two high-energy particles having similar energy can also be distinguished using this detector.

\begin{figure*}
  \centering
   {{\includegraphics[width=6.2cm]{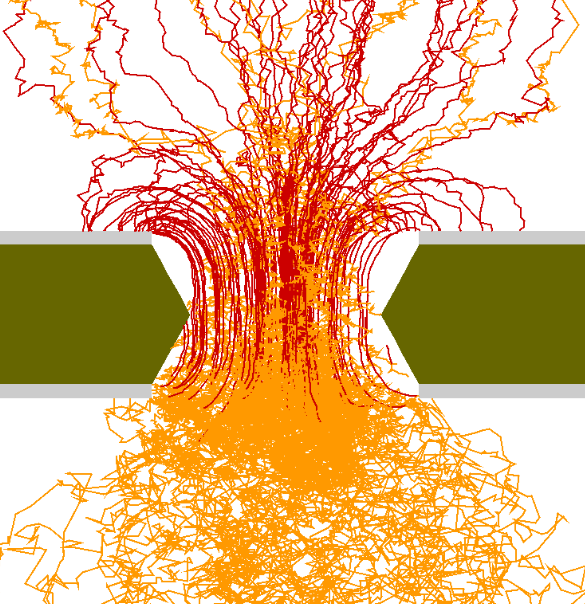} }}
    \caption{(Color online) The high electric field inside the GEM hole triggers primary and secondary ionization of gas molecules, resulting in an electron avalanche. Electrons, shown by yellow lines, drift towards the anode, while the ions, represented by red lines, move in the opposite direction towards the cathode. This process amplifies the initial ionization signal, enhancing the detection sensitivity in particle tracking and radiation monitoring.}
    \label{fig:1}
\end{figure*}


\subsection{Ion backflow}
After the avalanche process is over, electrons move toward the anode and most of the ions are collected by the lower and upper copper layer of the foil. However, some of the ions move backward to the cathode through the drift region which causes the ion backflow. The backflow of ions causes in reduction in the overall current of the ion electron cycle which is called the back current. The drift velocity of the ions is usually very low compared to the electron drift velocity due to its mass. Therefore, in the process ions accumulate in the drift region creating disturbances in the process and distorting the electric field which further reduces the gain. In addition, the ions recombine with the primary electrons, thereby restricting further avalanches. The detector's efficiency increases with increasing gain and therefore depends on the collected electrons on the anode plate which are produced from the avalanche. With higher ion backflow, the detector efficiency decreases subsequently. Higher back current led to several difficulties like increasing electric noise and decreasing the accuracy in detecting particle position and energy. Our main aim of this study is to design the GEM foil in such a way that, it increases the gain and reduces the ion backflow inside the detector to improve the performance and efficiency.

\section{Construction of GEM foil using ANSYS}

The gain and efficiency of the GEM detector depend on the geometry of the foil. In the standard GEM foil, 5 \(\mu\)m copper is cladded on both surfaces of 50 \(\mu\)m kapton. The pitch; center-to-center distance between two consecutive holes is 140 \(\mu\)m. The outer hole diameter is 70 \(\mu\)m and the inner hole diameter is 50 \(\mu\)m. The length of the drift region is 3 mm and the induction region is 2 mm.

\vspace{1em}
$\boldsymbol{Ansys: }$  Ansys is a finite element method (FEM) \cite{huebner2001finite} software that allows us to build the geometry of the gem detector and create electric field maps. Firstly, the full GEM detector geometry is built using Ansys Maxwell software \cite{martyanov2014ansys}. Then properties were assigned to the geometry and voltages were applied to the cathode, upper copper layer, lower copper layer, and anode. The strength of the average electric field inside the drift region, gem hole, and induction region was calculated.

The electric field strength \(E_{\text{strength}}\) is,
\begin{equation}
E_{\text{strength}} = \frac{\int E \, dV}{\int dV}
\label{eq:E_strength}
\end{equation}

\begin{figure*}
  \centering
   [$\mathsf{(a)}$]{{\includegraphics[width=3.2cm]{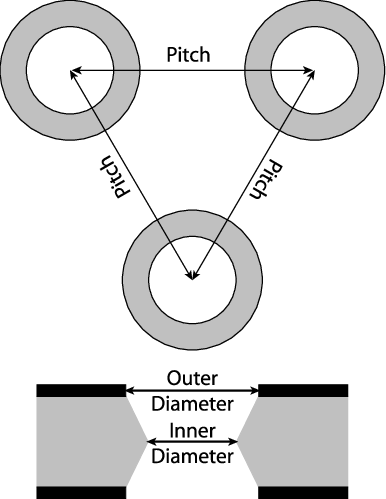} }}
   [$\mathsf{(b)}$]{{\includegraphics[width=6.4cm]{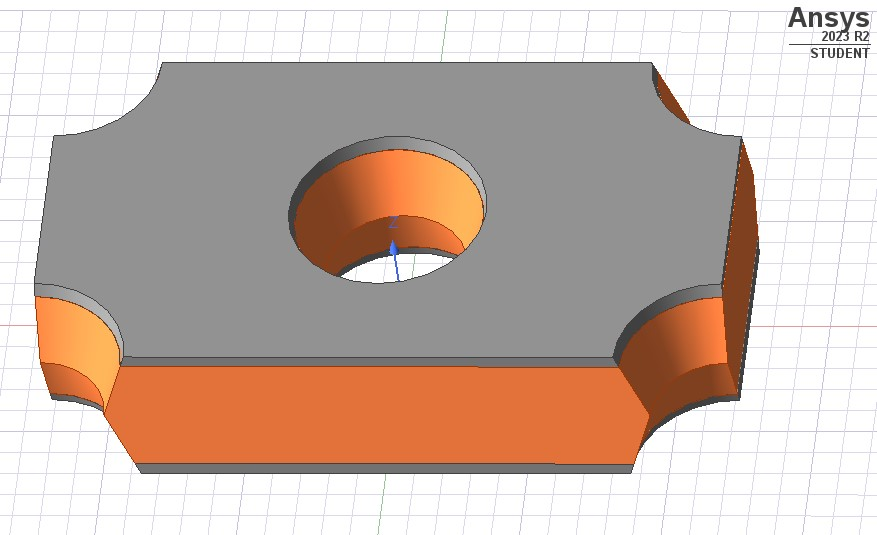} }}
   [$\mathsf{(c)}$]{{\includegraphics[width=6.0cm]{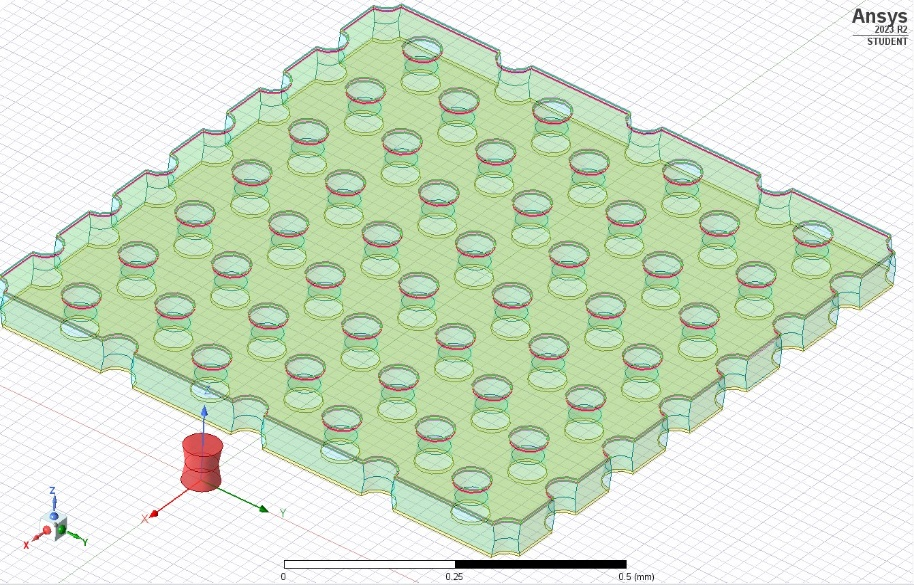} }}
    \caption{(color online) Illustrations of various aspects of GEM foil and detector geometry created using ANSYS Maxwell. (a) Illustration of pitch size which is the center-to-center distance between two holes of the gem foil. The inner and outer hole diameters of the GEM are also shown., (b) Unit GEM created in ANSYS Maxwell with inner hole diameter 50 \(\mu\)m, outer hole diameter 70\(\mu\)m, kapton thickness 50\(\mu\)m which is sandwiched between two copper layers of thickness 5 \(\mu\)m., (c) The GEM foil was created in ANSYS Maxwell by stacking many unit cells of GEM with pitch size 140 \(\mu\)m.}
    \label{fig:1}
\end{figure*}

\par
A unit cell of GEM is modeled in Ansys with appropriate dimensions. Materials are assigned to the cell. Then the copies of the cell are made and merged to make the GEM foil of the desired dimension. The foil is kept in a chamber of gases with a specific ratio. The potential is assigned to the copper layers and the faces of the gas chamber which act as an anode and cathode. The electric field in the different regions of GEM is calculated with ANSYS field calculator. The electric field plays a crucial role in the gain of the detector. A higher electric field results in a high acceleration of electrons which affects the electron avalanche process. Through scripting, a unit gem model was constructed using Ansys mechanical APDL \cite{thompson2017ansys}. The unit cell also includes the induction and drift regions. We store the field maps in .lis files. Within the map, electric field information is available at every single point within the GEM geometry, which is defined. In this study, we changed the geometry of the GEM hole and pitch size to achieve a higher gain and reduce the backflow of ions.



Mechanical APDL (Ansys parametric design language) is utilized for further detailed study. The ability to precisely apply voltage boundary conditions and automate the study process, including meshing, solving for electrostatic fields, and post-processing results, is made possible by APDL's scripting features. To maximize electron multiplication efficiency in particle detection applications, precise analysis and optimization of the electric field distribution within and around the GEM foil are made possible by precise control. In APDL, the geometry of GEM foil is made from the script then the voltage and material properties are assigned which generates the files that are used in Garfield$^{++}$ to simulate the gaseous ionization of electrons inside the GEM foil.

\begin{figure*}
  \centering
   [$\mathsf{(a)}$]{{\includegraphics[width=7.2cm]{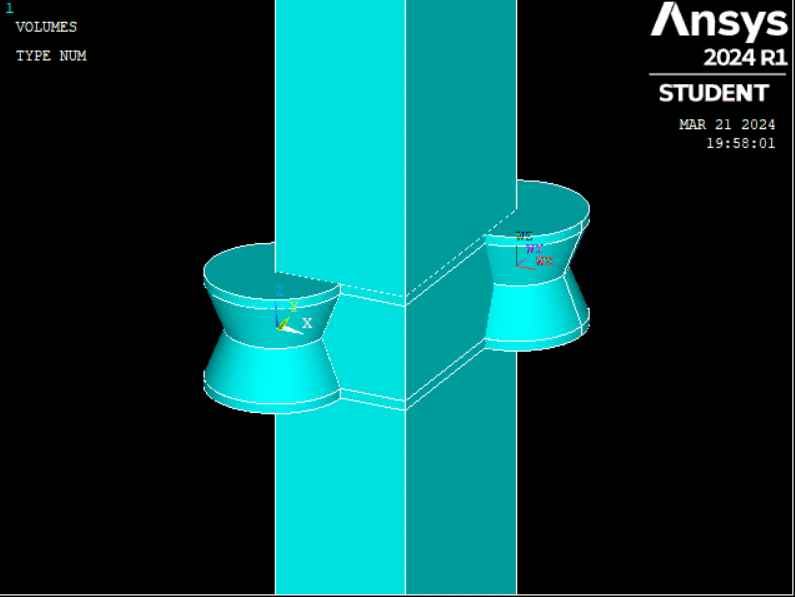} }}
   [$\mathsf{(b)}$]{{\includegraphics[width=7.2cm]{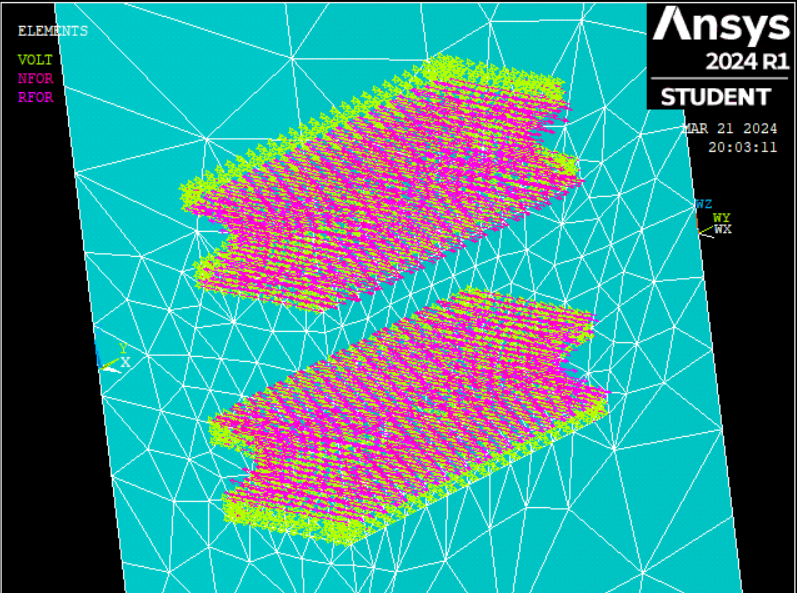} }}
    \caption{(color online) (a) GEM model in mechanical APDL (Ansys parametric design language) with double conical hole configuration (b) Meshing analysis of GEM unit in mechanical APDL.}
    \label{fig:1}
\end{figure*}

\vspace{0.5 cm}
\textbf{ELIST} (Element List): This file includes the geometrical elements that form the detector's structure. Each entry details the type, dimensions, and position of an element.
\newline
\textbf{NLIST} (Node List): This file contains the nodes in the mesh used for finite element analysis. Nodes are the connection points of the elements and the nlist specifies their coordinates.
\newline
\textbf{MPLIST} (Material Properties List): This file outlines the material properties for various regions or elements of the detector. It encompasses parameters such as permittivity, permeability, conductivity, and other specific material attributes.
\newline
\textbf{PRSNOL} (Potential and Response Node List): This file designates nodes where electric potential or response functions are calculated, aiding in the analysis of the detector's performance under electric fields and particle interactions.
The geometry of GEM created using APDL is shown in Figure 5.


\section{ study using Garfield$^{++}$}
Generated files created in ANSYS Mechanical APDL (.lis files) are used in Garfield$^{++}$\cite{sipaj2012simulation} for further study.
Garfield$^{++}$ is a powerful study tool designed for modeling gas-based detectors such as the Gas Electron Multiplier (GEM). It excels at simulating the complex behaviour of electrons and ions within the gas, accounting for the effects of electric fields, as well as the dynamics of charge multiplication and collection. By using advanced computational methods and integrating smoothly with tools like ANSYS and Magboltz, Garfield$^{++}$ delivers a thorough and accurate depiction of detector performance. With the power of C++, it efficiently manages complex calculations and large datasets, ensuring precise studies. Researchers can configure detailed geometries, specify material properties, and simulate particle interactions, making Garfield$^{++}$ an essential tool for optimizing GEM detector designs and improving their real-world functionality. This framework is designed for the in-depth analysis and modeling of particle detectors based on gas. Garfield$^{++}$ precisely models the process of electron avalanche by solving the equations of motion for electrons and ions, taking into account the influence of the electric field, gas composition, and the geometry of the detector. The \emph{.lis} (\emph{ELIST.lis, NLIST.lis, MPLIST.lis, PRSNOL.lis}) files are imported in Garfield$^{++}$ to simulate the behaviour of ions and electrons. The study includes stochastic processes such as electron collisions, energy loss, and secondary ionization events providing a detailed and accurate representation of the avalanche phenomenon. This enabled us to predict the performance of gas detectors, optimize their design, and better understand the underlying physical processes. Garfield$^{++}$  study of electron avalanche is shown in Figure 3. Garfield$^{++}$ It interfaces with Magboltz to calculate electron transport through gas using Monte Carlo study. By numerically integrating the Boltzmann transport equation, i.e., simulating an electron bouncing around inside a gas, the Magboltz program can calculate the properties of drift gases. It can calculate the drift velocity, starting point, and end point of ions and electrons under the influence of electric and magnetic fields by tracking the virtual electron propagation distance.. Yellow and red lines denote the drift of electrons and ions respectively. An electron with a lower ionization energy of e0=0.1 eV is used to simulate the GEM in Garfield$^{++}$. The initial position of the electron is randomly selected. After the avalanche, the endpoints of the electrons and ions are calculated. If the endpoints of the ions are above 50 \(\mu\)m from the upper copper foil, these ions are considered to be the backflow of ions. The number of electrons created in the avalanche process for single-line spectra is referred to as the gain. In Garfield$^{++}$, the study is conducted for 10 events of single-line spectra.

\section{Results}
Several studies have been performed to increase the performance of the GEM detector with different foil geometries. Detail results on the variations of the GEM foil's outer hole diameter and how it affects the detector's gain and performance are shown and discussed. Additionally, a novel hole geometry was implemented to attain increased gain. Studies have been performed by modifying the copper thickness of the GEM foil's lower side to attract the created ions which reduces the ion backflow. Finally, result have been shown to compare the ratio of ion backflow to the gain of the new types of geometry with the existing standard configuration.

\subsection{Effect of changing outer hole diameter}
Keeping the inner hole diameter constant, the outer hole diameter was changed, and the gain was calculated using Garfield$^{++}$ software for a single electron with an initial energy 0.1 eV. The study is done for 10 events. The quantity of primary electrons that could penetrate through a hole was small for a small outer hole diameter. In this case, most of the electrons were obstructed by the copper layer. In Figure 6(a), a GEM hole with an inner diameter of 50 \(\mu\)m and an outer diameter of 30 \(\mu\)m was studied using Garfield$^{++}$.

\begin{figure*}
  \centering
   [$\mathsf{(a)}$]{{\includegraphics[width=8.2cm]{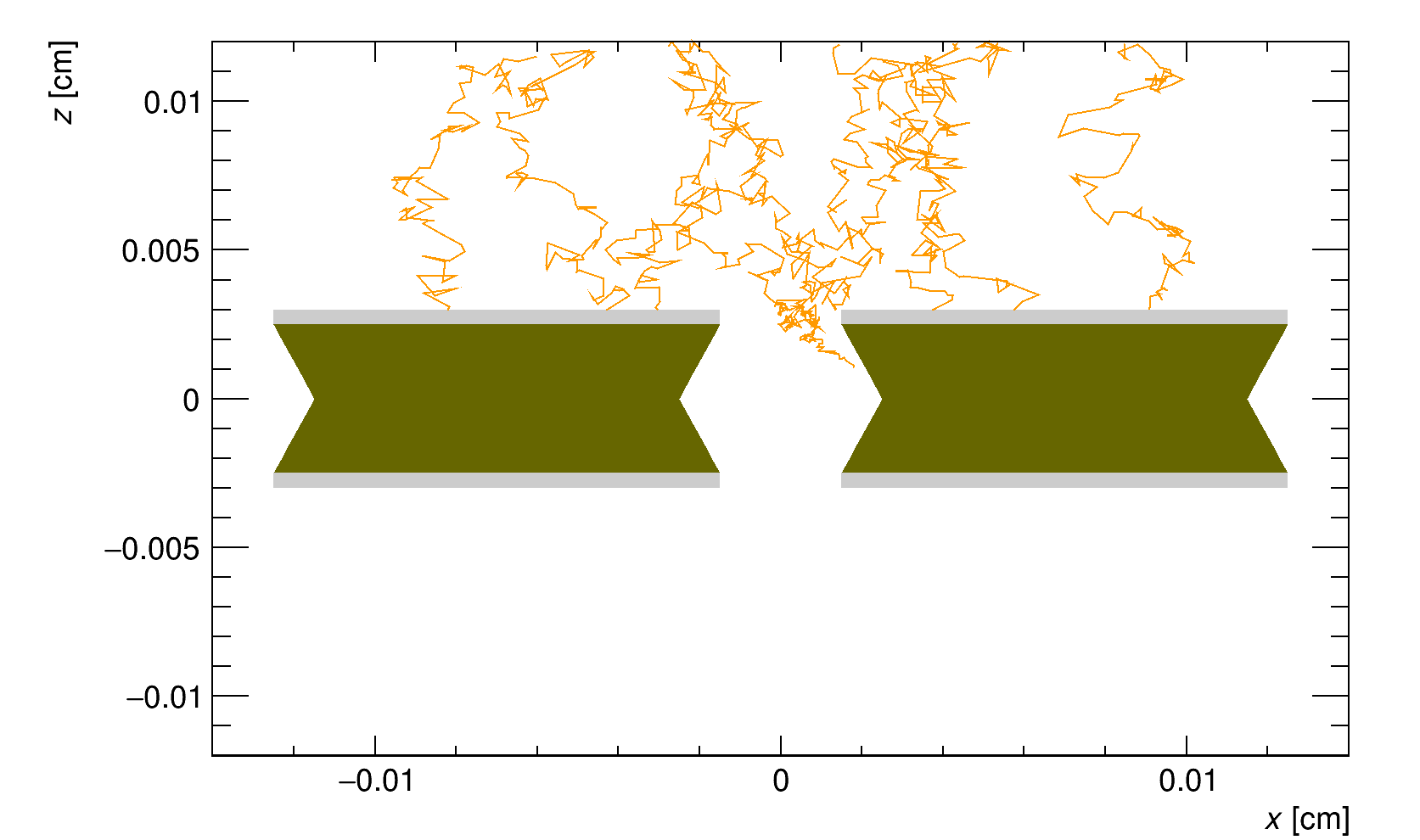} }}
   [$\mathsf{(b)}$]{{\includegraphics[width=8.2cm]{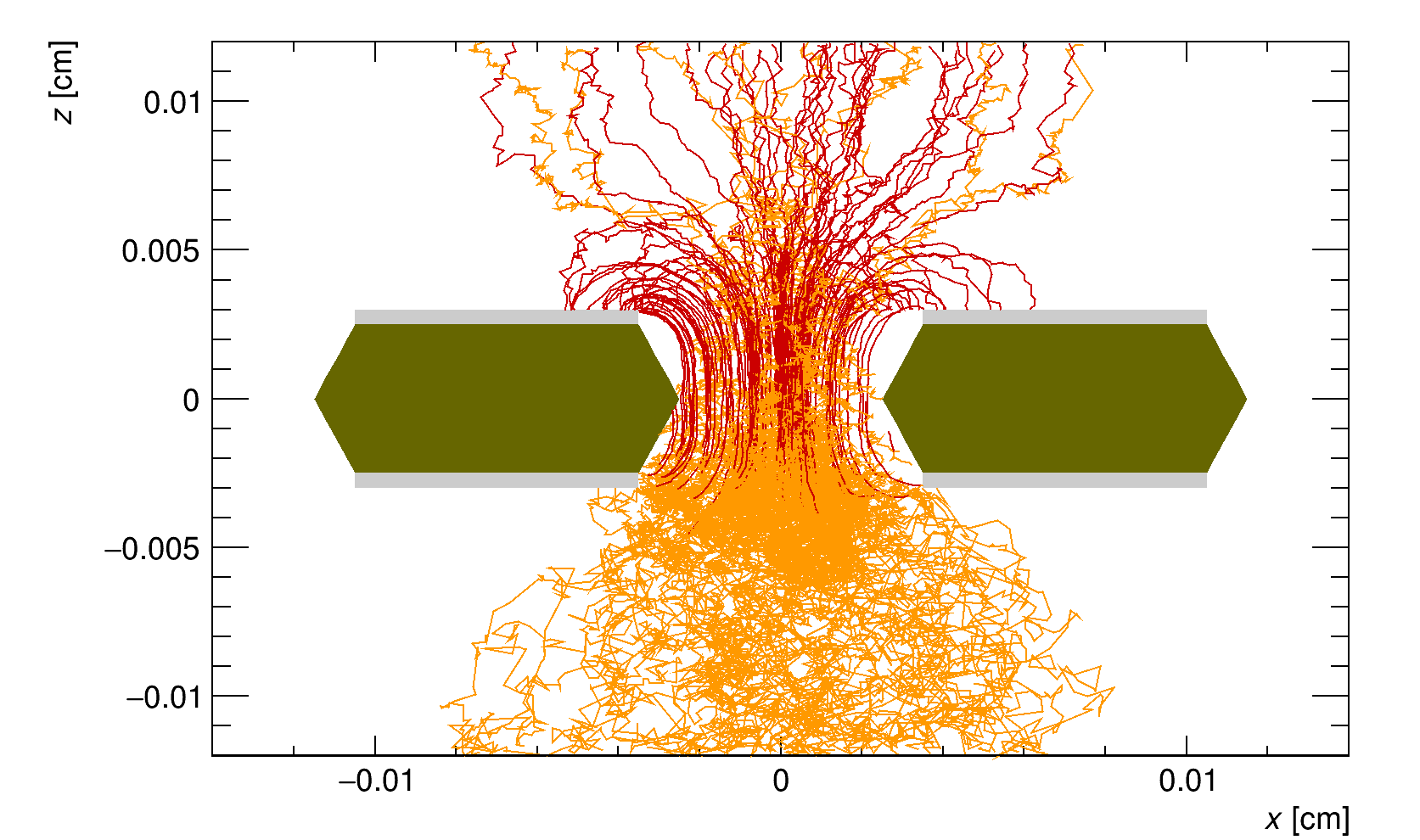} }}
   [$\mathsf{(c)}$]{{\includegraphics[width=8.2cm]{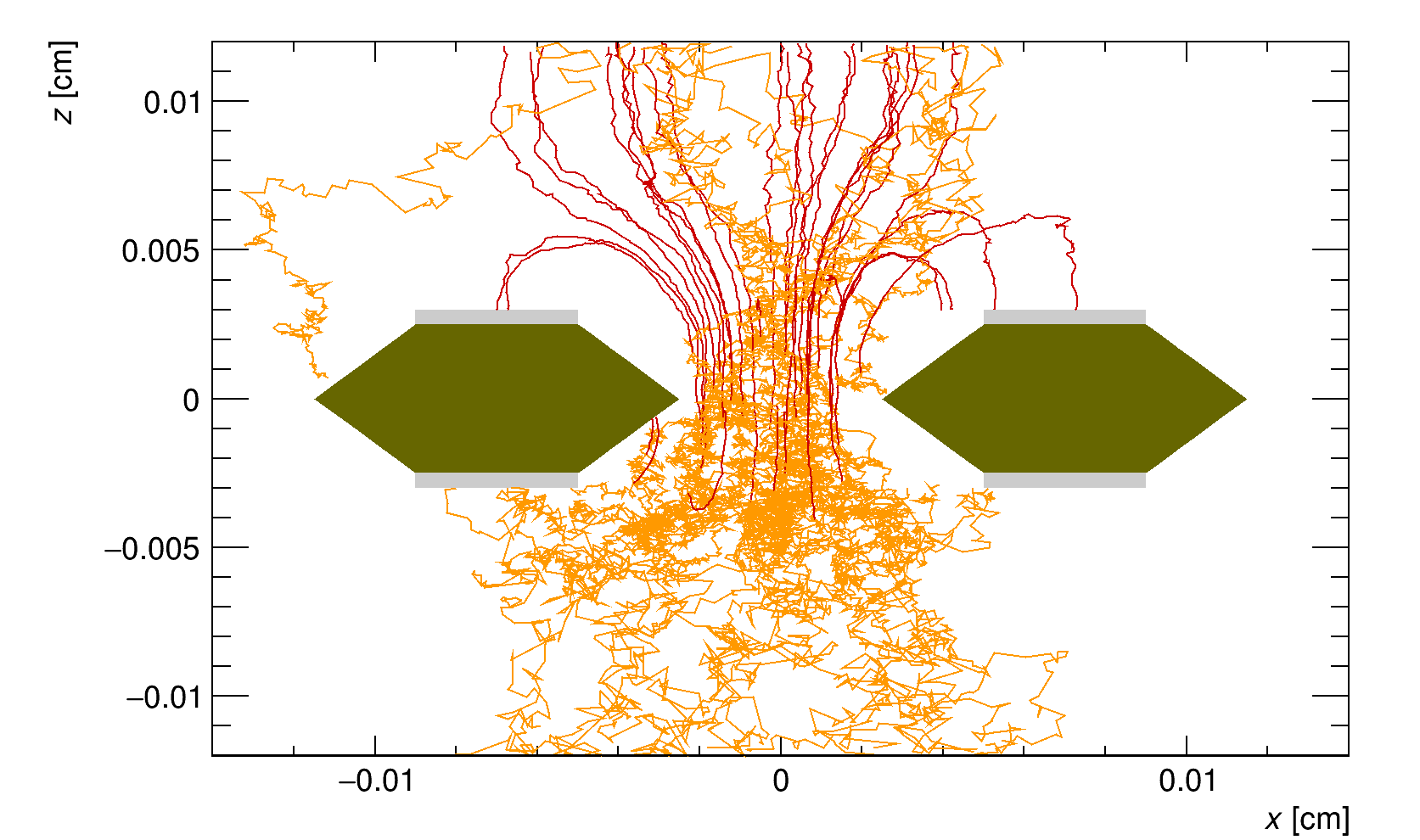} }}
    \caption{(color online) This graphical study of drift lines for the particles was created using Garfield$^{++}$. The yellow lines indicate the drift lines of electrons, while the red lines indicate the drift lines of ions. (a) define the study for inner GEM hole diameter 50 \(\mu\)m and outer diameter 30 \(\mu\)m, (b) define the study for inner diameter 50 \(\mu\)m and outer diameter 70 \(\mu\)m, (c) define the study for inner diameter 50 \(\mu\)m and outer diameter 100 \(\mu\)m.}
    \label{fig:1}
\end{figure*}

\begin{figure*}
  \centering
   {{\includegraphics[width=10.2cm]{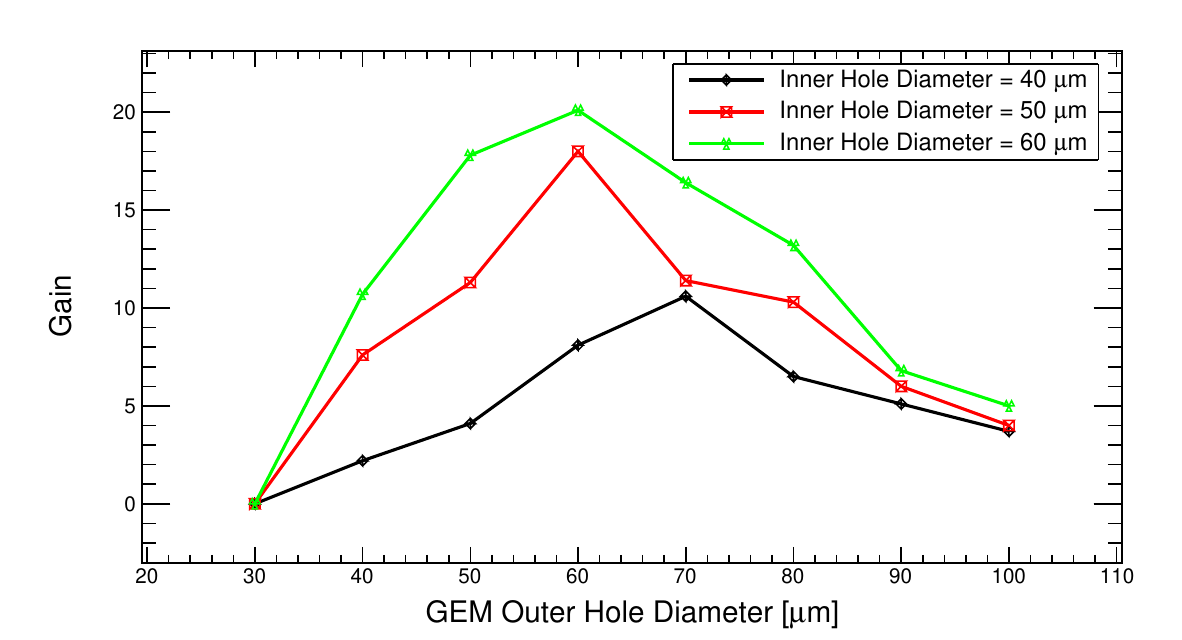} }}
    \caption{(color online) Gain of GEM for various outer hole diameters with fixed inner hole diameters. Green, red, and black lines denote the inner hole diameter of 60 \(\mu\)m, 50 \(\mu\)m, and 40 \(\mu\)m respectively.}
    \label{fig:1}
\end{figure*}

No electrons could penetrate due to the smaller outer hole diameter. More electrons could penetrate through a larger outer hole diameter, but the average electric field inside the hole was minimal. According to Eq~\ref{eq:E_strength} as the volume increased, the average electric field inside the hole should decrease. Consequently, the avalanches inside the holes were limited, and the gain was small. In Figure 6(c), GEM hole with an inner diameter of 50 \(\mu\)m and an outer diameter of 100 \(\mu\)m was studied using Garfield$^{++}$. Most of the electrons penetrated through the hole, but the avalanche was minimal due to the lower electric field strength. For both smaller and larger outer hole diameters, the gain was small. On the other hand, the highest gain was observed when the outer hole diameter approached the thickness of the Kapton foil. This condition corresponded to the maximum avalanche occurrence and the highest value of the electric field. In Figure 6(b), GEM hole with an inner diameter of 50 \(\mu\)m and an outer diameter of 70 \(\mu\)m was studied using Garfield$^{++}$. In this case, the gain was maximum. Figure 7 shows the variation of the gain for different inner hole diameters. The black, red, and green lines indicate fixed inner hole diameters of 40 \(\mu\)m, 50 \(\mu\)m, and 60 \(\mu\)m, respectively. The gain increased with the increase of the outer hole diameter. When the outer hole diameter was between 50 \(\mu\)m and 70 \(\mu\)m, the gain attains its maximum value. After the maximum value of the gain decreases with increasing outer hole diameter.

\begin{figure*}
  \centering
   [$\mathsf{(a)}$]{{\includegraphics[width=4.8cm]{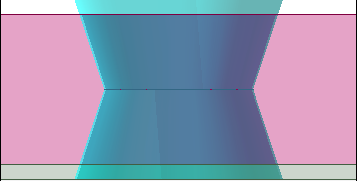} }}
   [$\mathsf{(b)}$]{{\includegraphics[width=4.8cm]{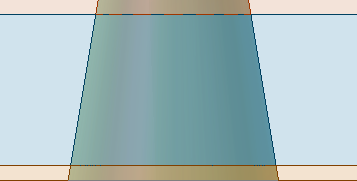} }}
   [$\mathsf{(c)}$]{{\includegraphics[width=4.2cm]{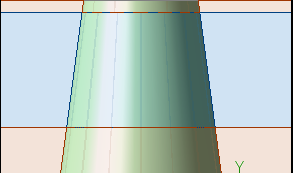} }}
    \caption{(color online) (a) Represents the bi-conical shaped hole geometry of inner hole diameter 50 \(\mu\)m and outer hole diameter 70 \(\mu\)m with upper and lower copper thickness of 5 \(\mu\)m, (b) Represents the single conical hole geometry with an upper diameter of 50 \(\mu\)m and a lower diameter of 70 \(\mu\)m, featuring copper thicknesses of 5 \(\mu\)m on both the upper and lower sides of the GEM foil, (c) Single conical shaped hole with the upper copper thickness 5 \(\mu\)m and lower copper thickness 15 \(\mu\)m.}
    \label{fig:1}
\end{figure*}

\subsection{Single conical shape hole}
The existing geometry of GEM has an inner diameter of 50 \(\mu\)m and an outer diameter of 70 \(\mu\)m. When a highly energetic charged particle entered the drift region, it generated primary electrons and ions. The electric field pulled electrons toward the GEM foil. Due to the intense electric field within the GEM's holes, an avalanche ensued. Created electrons moved toward the anode, while the ions moved backward due to the electric field. As the ions moved backward, some of them were trapped on the lower side of the Kapton foil, unable to move due to the conical shape. As events occurred, the number of trapped ions increased. This ion accumulation on the Kapton foil narrowed the path for electrons traveling through the holes, reducing the number of electrons that could penetrate the hole. Consequently, the number of avalanches decreased due to this obstruction, affecting the detector's performance. To counter this problem, a new type of hole geometry, a single conical-shaped hole, was introduced. A single conical-shaped hole with an upper diameter of 50 \(\mu\)m and a lower diameter of 70 \(\mu\)m was used in the GEM foil for electron avalanche. For the single conical-shaped hole, after the avalanche, the ions were trapped on the upper side of the hole. Thus, the single conical-shaped hole provided a larger volume on the upper side compared to the lower side of the existing conical hole, allowing more space for further avalanches. The single conical-shaped geometry resulted in a higher avalanche rate than the existing conical shape. Consequently, the efficiency of the single conical-shaped hole increased compared to the existing geometry. In Figures 8(a) and 8(b), a biconical and a single conical-shaped hole are shown which are created using ANSYS Maxwell. A detailed study of the gain of these foils with new types of hole geometries is performed. The gains for existing standard GEM and newly modeled single conical-shaped hole GEM are compared and were found to be 15.6 and 23.0, respectively, in the 1-line spectra.

\subsection{Variation of the lower copper thickness}

To achieve high gain, the hole geometry was changed from biconical to single conical-shaped. However, for the single conical-shaped hole, a problem has been raised that the upper hole diameter is 50 \(\mu\)m and the lower hole diameter is 70 \(\mu\)m, making the hole size uneven between the upper and lower sides of the GEM foil. The single conical-shaped GEM hole shown in Figure 8(b) with an upper hole diameter of 50 \(\mu\)m and a lower hole diameter of 70 \(\mu\)m has a similar copper thickness as in the standard GEM foils. On the lower side of the foil, there is a larger open space than on the upper side, making the foil very fragile and difficult to handle. To handle these problems, the lower copper thickness was increased slowly to attract the trapped ions and to provide additional steadiness. Figure 8(c) shows a single conical-shaped GEM hole geometry with an upper copper thickness of 5 \(\mu\)m and a lower copper thickness of 15 \(\mu\)m. The avalanche inside the hole occurs on the lower side. Ions created during the avalanche process are collected by both the lower and upper copper layers. With the increase in the lower copper thickness, more ions can be collected by the copper, potentially reducing ion backflow. Due to the change in copper thickness, the electric field in each region of the detector is altered, which affects the gain and the ion backflow. To lower the back current, the ion back-flow also needs to be reduced. The gain is studied by considering two distinct cases.

\vspace{1em}
\par
$\boldsymbol{Case I: V_{GEM}\ constant:}$ When the potential difference between the copper layers of the GEM foil is kept constant $(\Delta V_{\text{GEM}} = 300 Volt)$,  the volume inside the hole increases with an increase in the lower copper thickness. As the electric field inside the hole geometry is inversely proportional to the volume (see Eq~\ref{eq:E_strength} ), the electric field is observed to decrease with an increase in copper thickness.

\begin{figure*}
  \centering
   [$\mathsf{(a)}$]{{\includegraphics[width=8.2cm]{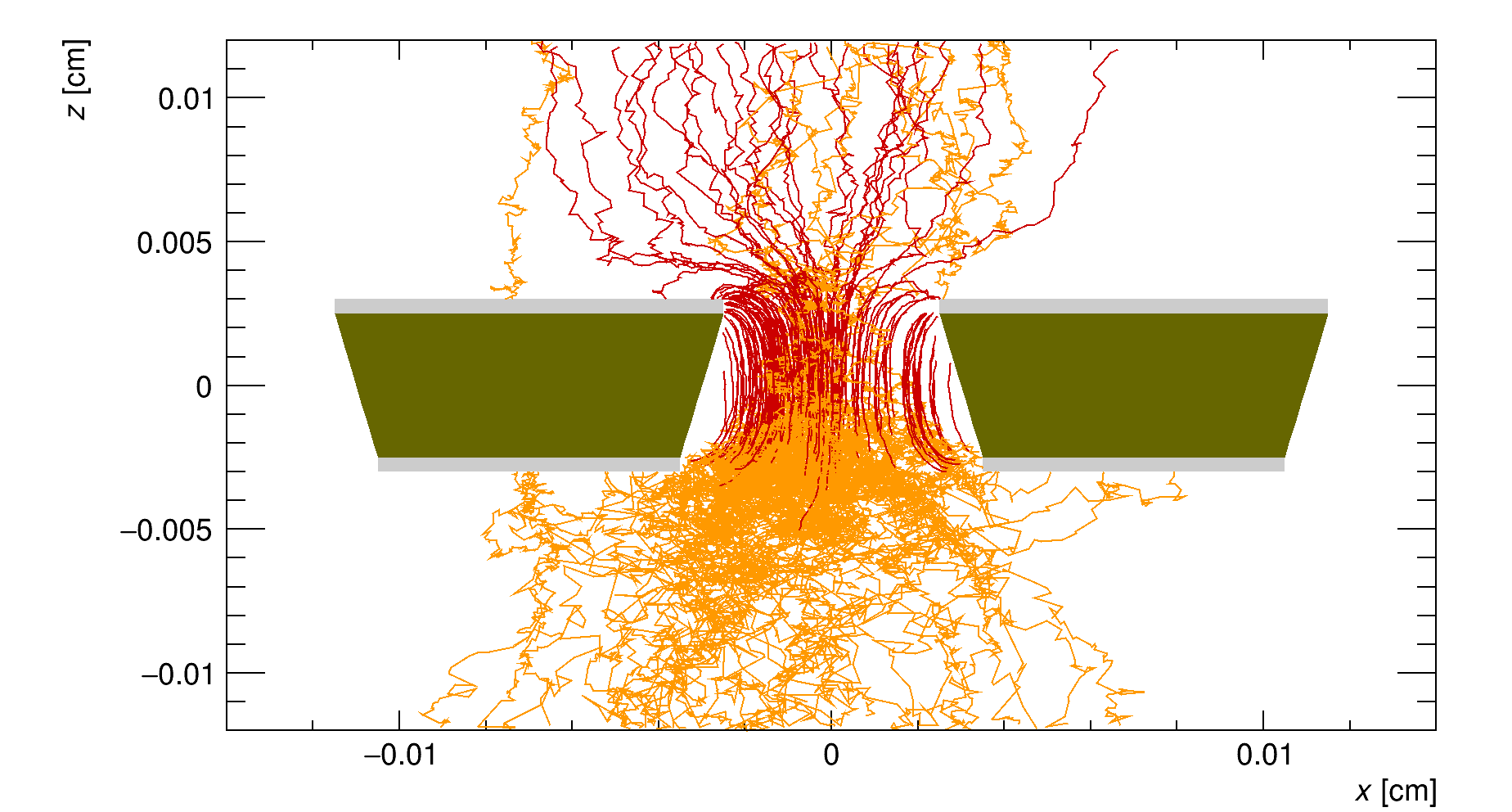} }}
   [$\mathsf{(b)}$]{{\includegraphics[width=8.2cm]{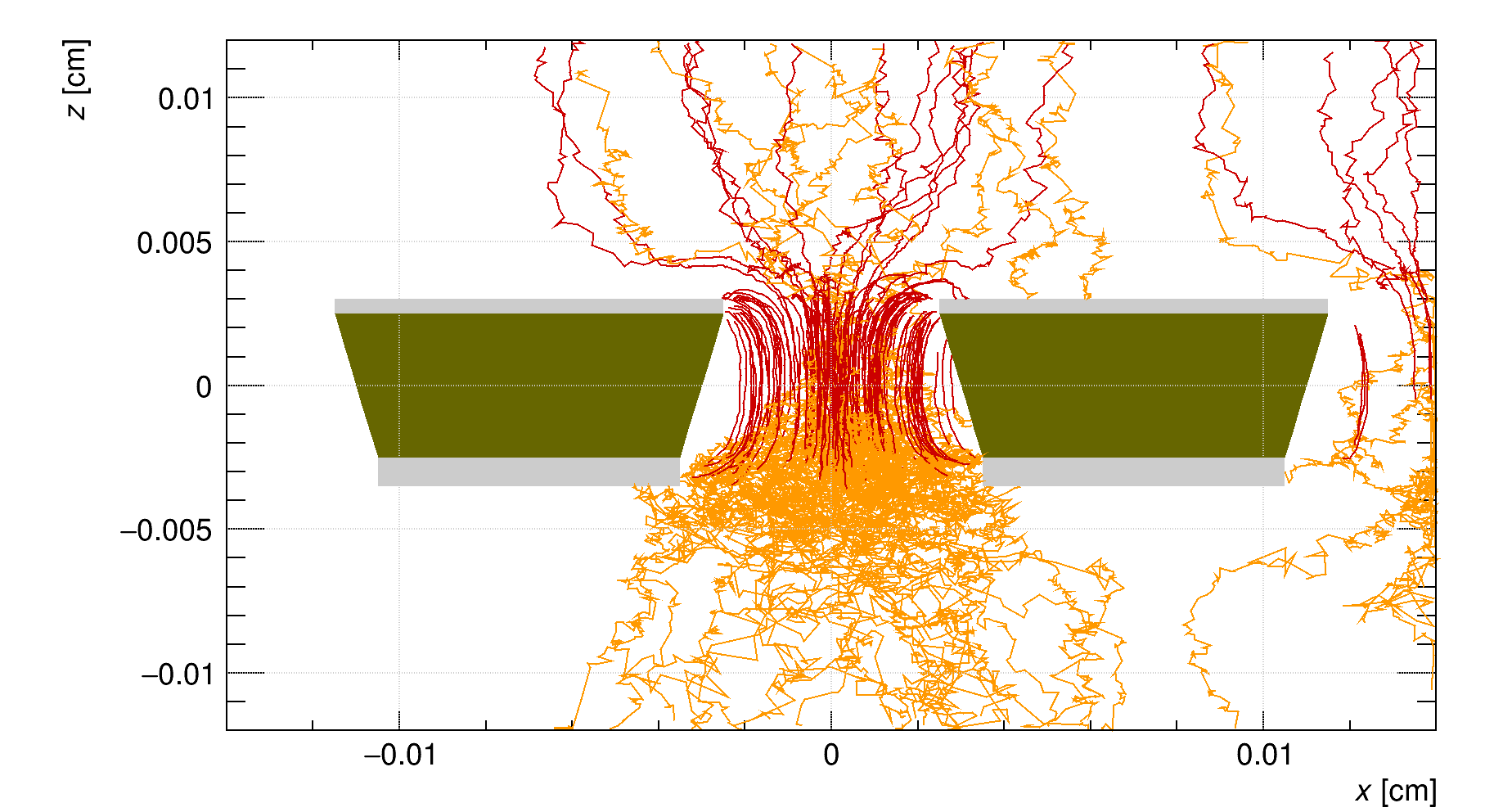} }}
   [$\mathsf{(c)}$]{{\includegraphics[width=8.2cm]{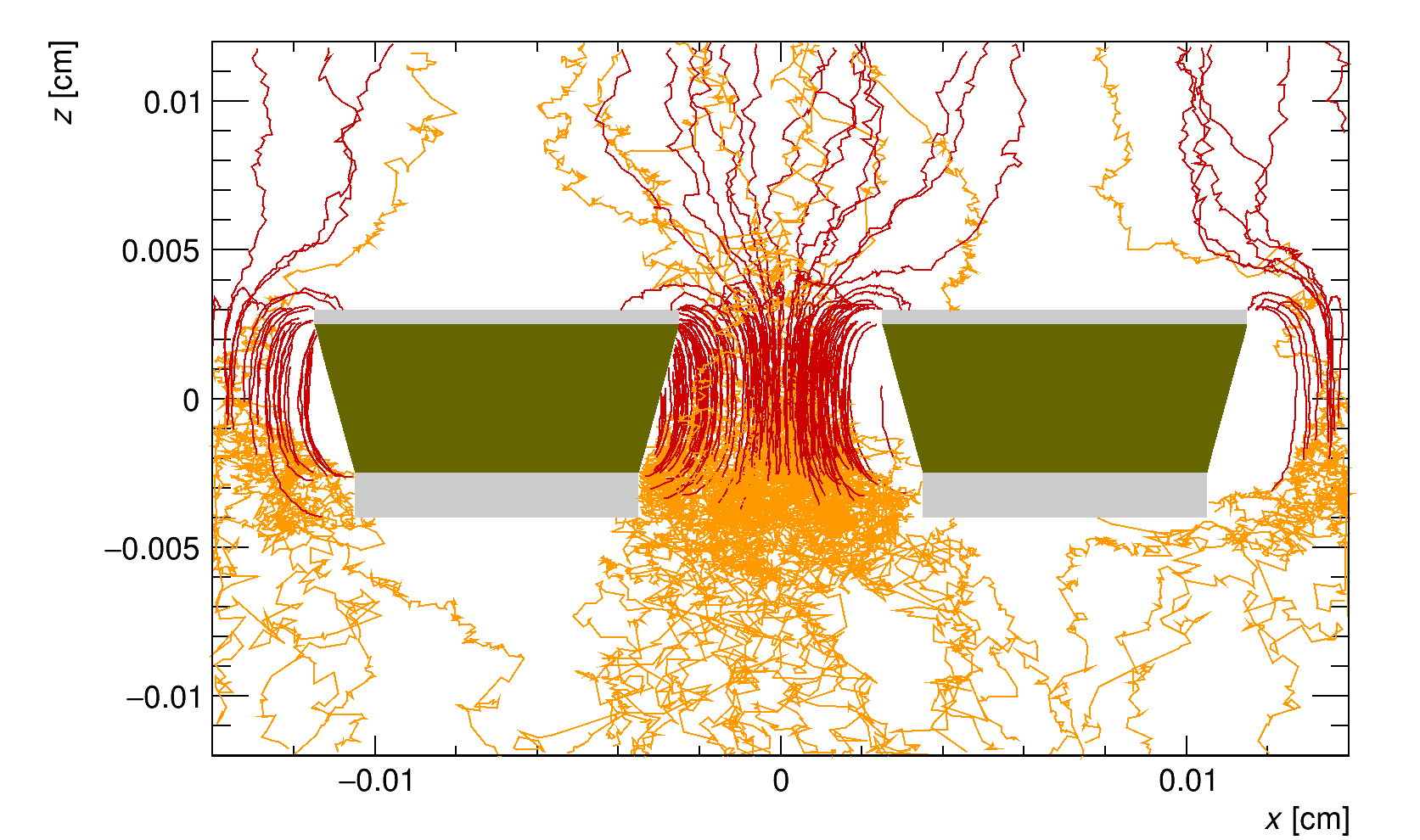} }}
   [$\mathsf{(d)}$]{{\includegraphics[width=8.2cm]{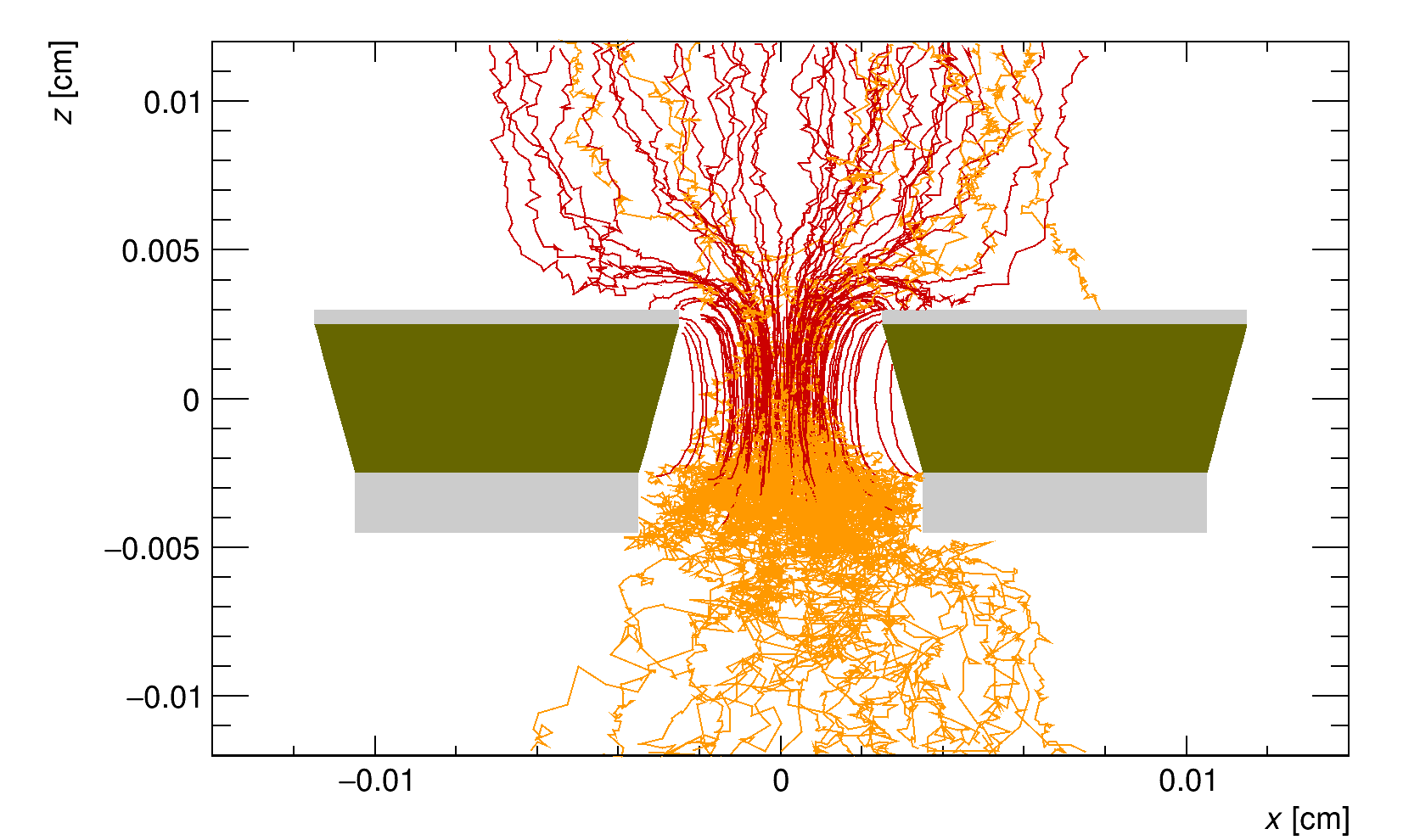} }}
    \caption{(color online) Using Garfield$^{++}$ software, an electron avalanche was simulated inside a single conical GEM hole with an upper diameter of 50 \(\mu\)m and a lower diameter of 70 \(\mu\)m for different lower side copper thicknesses. During the study, the potential difference between the copper layers of the GEM foil was fixed. Consequently, the electric field inside the hole decreased as the copper thickness increased. As the lower copper thickness increased, the avalanche inside the GEM hole decreased, reducing the gain of the detector. The study was conducted for 10 events. The yellow lines represent the drift lines of the electrons, and the red lines represent the drift lines of the ions. (a) Copper thickness of the upper side of the foil is 5 \(\mu\)m and copper thickness of the lower side of the foil is also 5 \(\mu\)m, (b) Copper thickness of the upper side of the foil is 5 \(\mu\)m and copper thickness of the lower side of the foil is also 10 \(\mu\)m, (c) Copper thickness of the upper side of the foil is 5 \(\mu\)m and copper thickness of the lower side of the foil is also 15 \(\mu\)m, (d) Copper thickness of the upper side of the foil is 5 \(\mu\)m and copper thickness of the lower side of the foil is also 20 \(\mu\)m.}
    \label{fig:1}
\end{figure*}

The gain of the detector depends on the electric field strength and the geometry of the foil. In Figure 9, the study conducted with Garfield$^{++}$ illustrates the avalanche occurring within a single conical GEM hole, along with the drift lines of ions and electrons. The hole has an upper diameter of 50 \(\mu\)m and a lower diameter of 70 \(\mu\)m, with a fixed electric potential between the two copper layers. The results show that the avalanche decreases as the thickness of the copper layer increases. The study was done for four different thicknesses of the lower copper layers, those are 5 \(\mu\)m, 10 \(\mu\)m, 15 \(\mu\)m, and 20 \(\mu\)m. According to the black curve in Figure 11(a), the gain is seen to slightly increase from a lower copper thickness of 2 \(\mu\)m to a copper thickness of 10 \(\mu\)m. After this point, the gain begins to decrease. After avalanches, the ions are collected by both copper layers. With an increase in the thickness of the lower copper, it absorbs more ions after the avalanche. As illustrated in Figure 11(b), the quantity of ions that flow back diminishes with an increase in copper thickness. The black lines represent this outcome.

$\boldsymbol{Case II: The\ electric\ field\ constant\ inside\ the\ regions:}$ In the previous case, to keep the potential difference unchanged between two copper layers of the GEM foil, the gain of the detector decreased. To achieve a higher gain the $V_{GEM}$ was adjusted to maintain a steady electric field in every region.  In Figure 10, the Garfield$^{++}$  study demonstrates the avalanche phenomenon within a single conical GEM hole, along with the drift lines of both ions and electrons. The GEM hole features an upper diameter of 50 \(\mu\)m and a lower diameter of 70 \(\mu\)m, with a consistent electric field throughout the detector. The findings indicate that as the thickness of the lower side of the copper layer increases, the avalanche also increases. The study was done for four different thicknesses of the lower copper layers, those are 5 \(\mu\)m, 10 \(\mu\)m, 15 \(\mu\)m, and 20 \(\mu\)m. Specifically, when the lower copper layer measures 5 \(\mu\)m, the avalanche is at its lowest, and then it starts to increase. The gain peaks when the thickness reaches 20 \(\mu\)m. According to the red line in Figure 11(a), the gain rises with increasing lower copper thickness. After the avalanches, both copper layers collect the ions, and a thicker lower copper layer can capture more ions, and the backflow of the ions can be reduced. Figure 11(b) shows that the number of ions flowing back remains constant as copper thickness increases, as indicated by the red lines. With a higher avalanche, the number of both electrons and ions increases. Thus, the constant line indicates that the majority of the ions produced are collected by the thicker copper layer.

\begin{figure*}
  \centering
   [$\mathsf{(a)}$]{{\includegraphics[width=8.2cm]{Scgemlc005.png} }}
   [$\mathsf{(b)}$]{{\includegraphics[width=8.2cm]{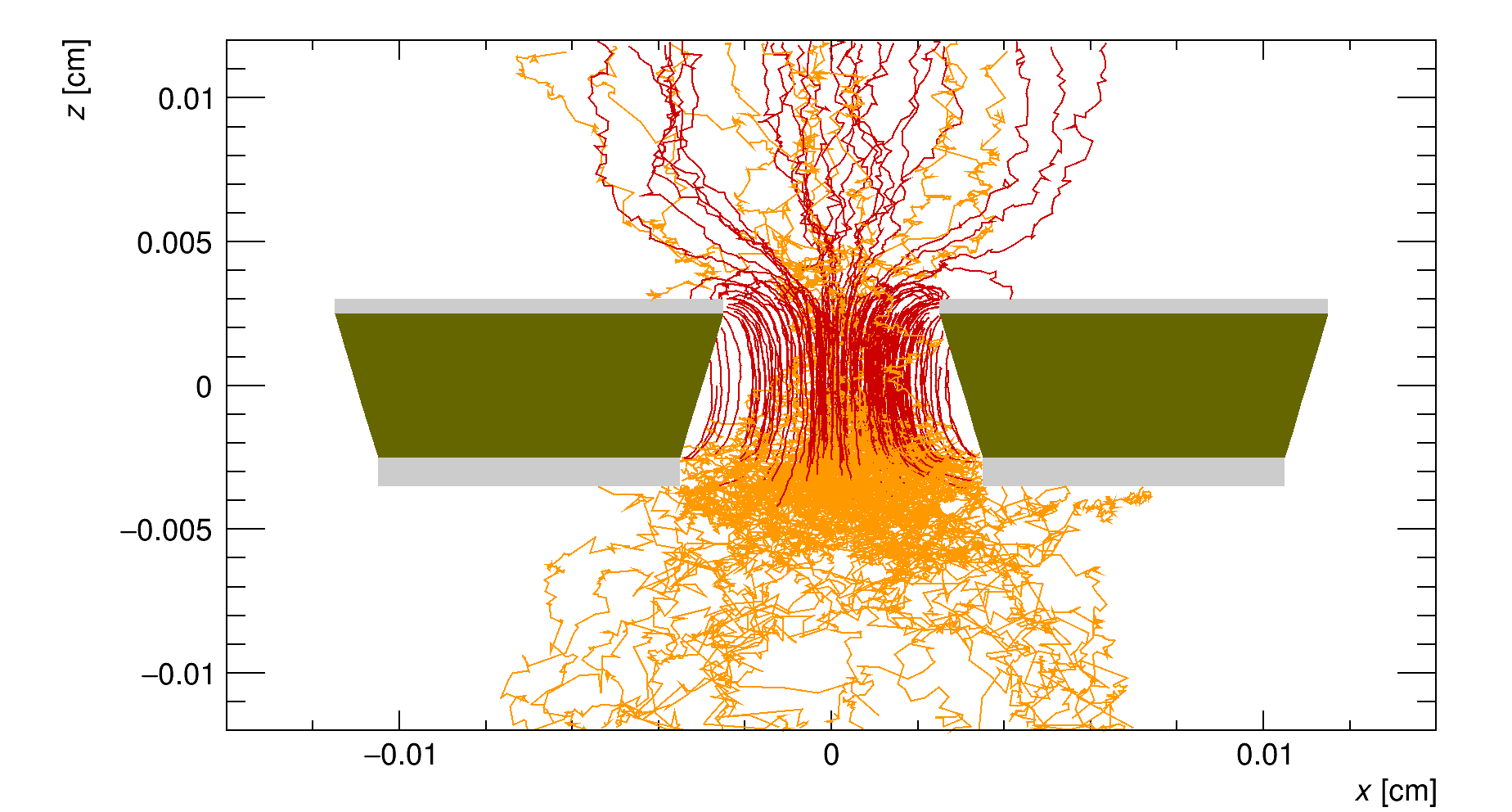} }}
   [$\mathsf{(c)}$]{{\includegraphics[width=8.2cm]{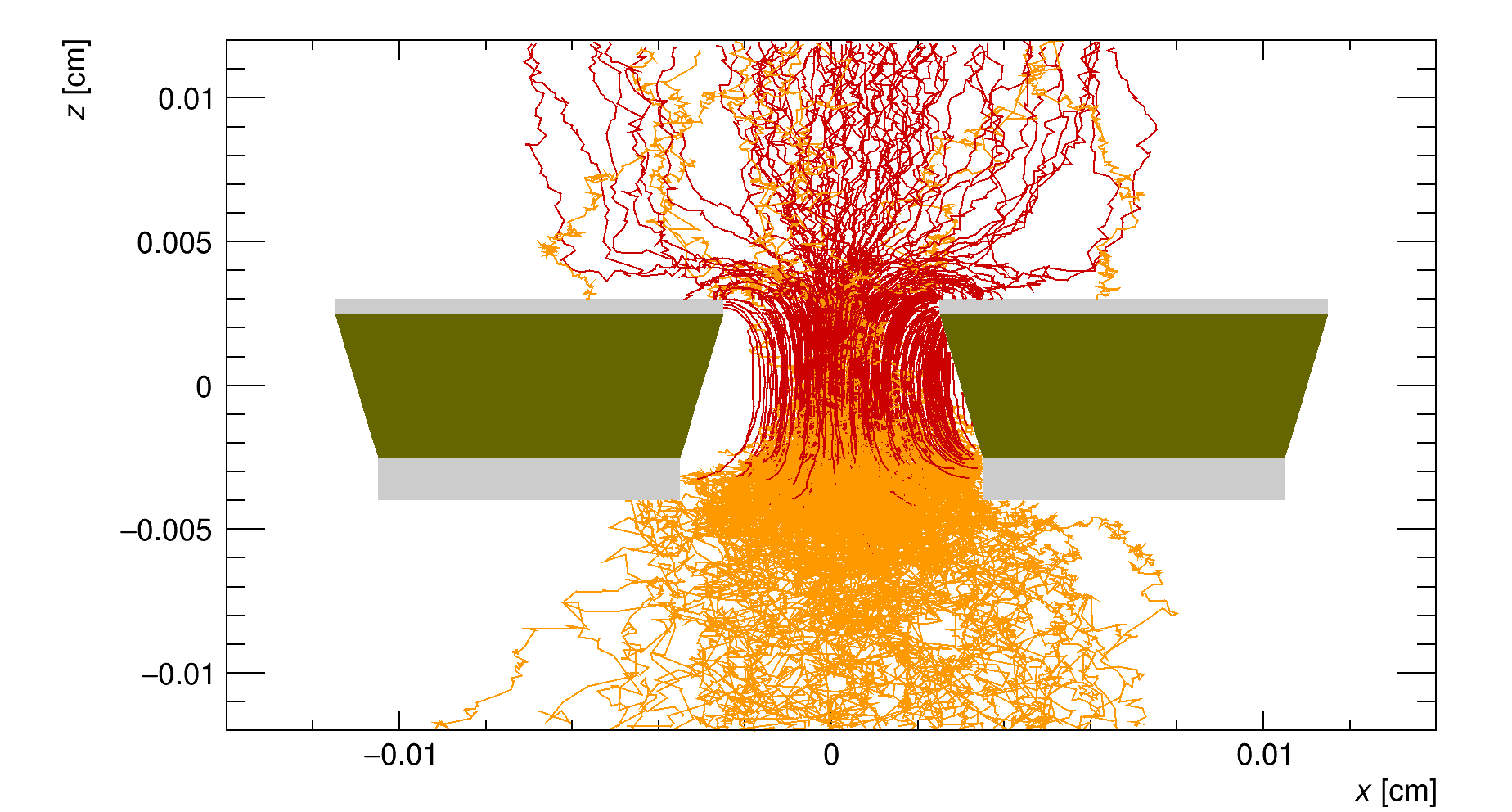} }}
   [$\mathsf{(d)}$]{{\includegraphics[width=8.2cm]{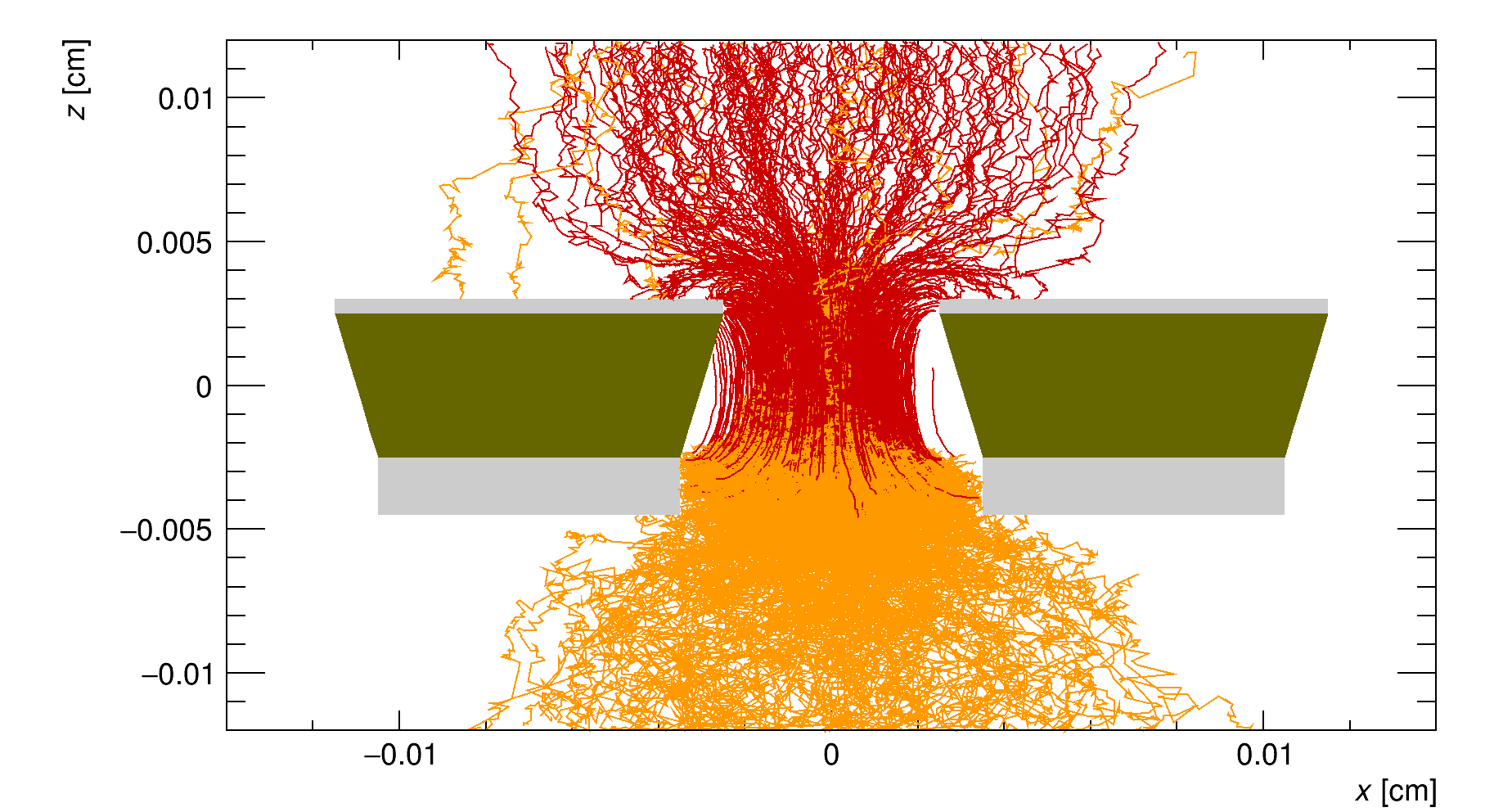} }}
    \caption{(color online) Using Garfield$^{++}$ software, an electron avalanche was simulated inside a single conical GEM hole with an upper diameter of 50 \(\mu\)m and a lower diameter of 70 \(\mu\)m for different lower side copper thicknesses. During the study, the electric field strength inside every region was fixed to change the potential difference between the copper layers. As the lower copper thickness increased, the avalanche inside the GEM hole increased, increasing the gain of the detector. The study was conducted for 10 events. The yellow lines represent the drift lines of the electrons, and the red lines represent the drift lines of the ions. (a) Copper thickness of the upper side of the foil is 5 \(\mu\)m and copper thickness of the lower side of the foil is also 5 \(\mu\)m, (b) Copper thickness of the upper side of the foil is 5 \(\mu\)m and copper thickness of the lower side of the foil is also 10 \(\mu\)m, (c) Copper thickness of the upper side of the foil is 5 \(\mu\)m and copper thickness of the lower side of the foil is also 15 \(\mu\)m, (d) Copper thickness of the upper side of the foil is 5 \(\mu\)m and copper thickness of the lower side of the foil is also 20 \(\mu\)m.}
    \label{fig:1}
\end{figure*}

 The efficiency of the detector depends on the ratio of ion backflow to gain. Ion backflow distorts the electric field inside the GEM hole, which further reduces the gain of the detector. Additionally, ions can recombine with primary electrons in the drift region, limiting further avalanches. A decrease in this ratio leads to an increase in the detector's efficiency. 
\begin{figure*}
  \centering
   [$\mathsf{(a)}$]{{\includegraphics[width=7.2cm]{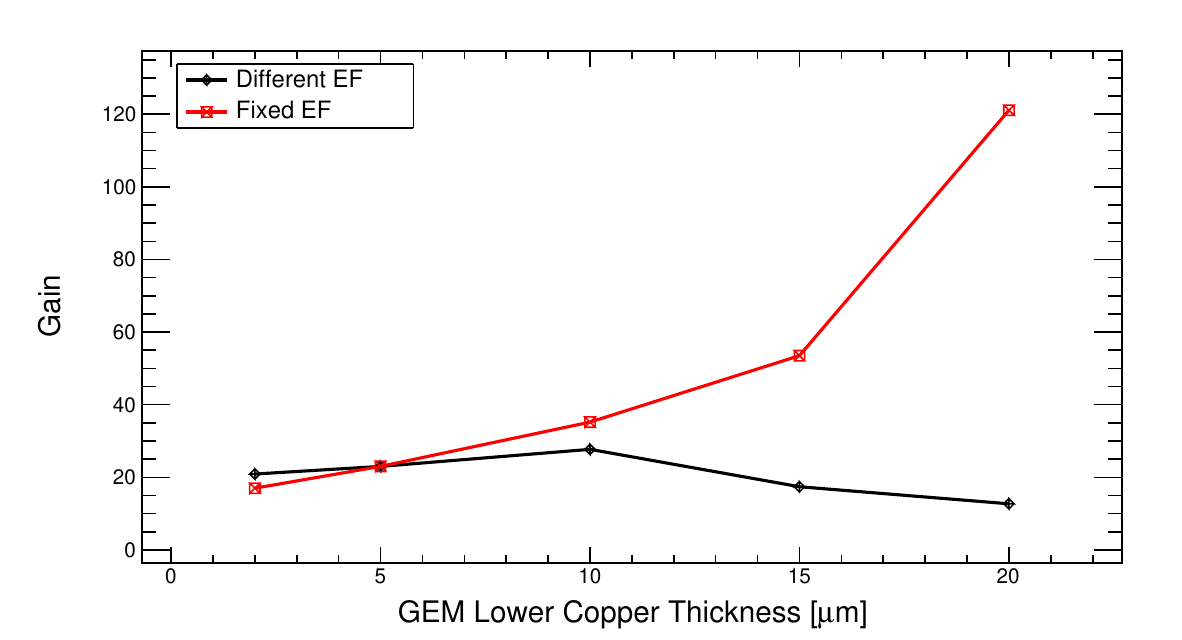} }}
   [$\mathsf{(b)}$]{{\includegraphics[width=7.2cm]{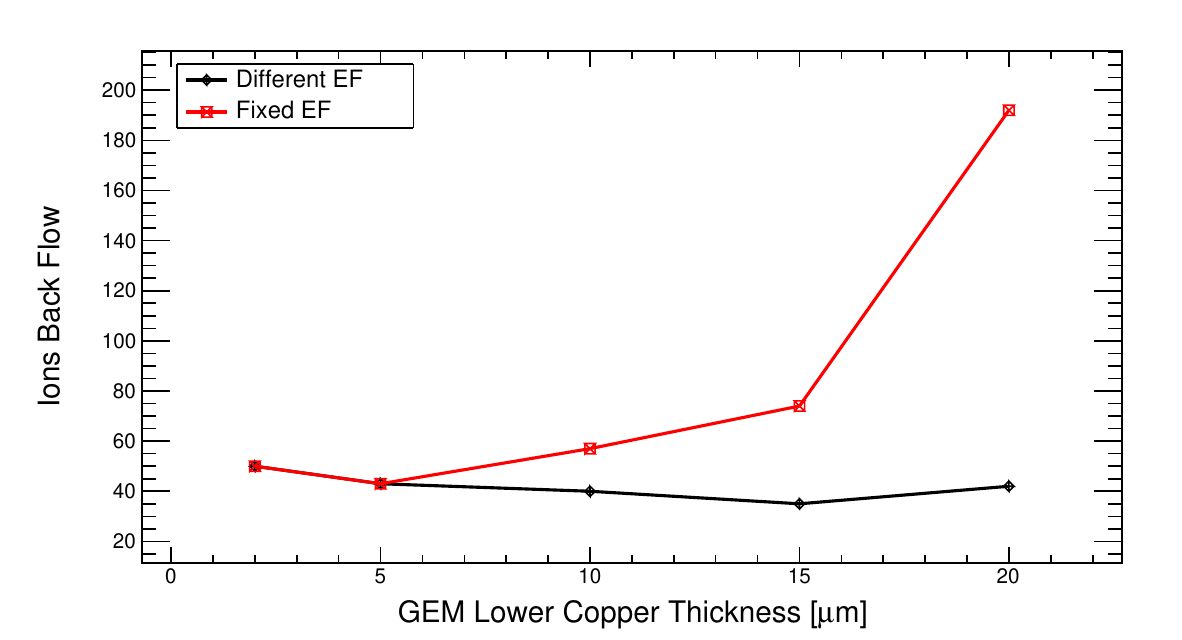} }}

    \caption{(color online)Studies conducted with variation in lower copper thickness using Garfield$^{++}$. In both plots, the black line indicates Case I: $V_{GEM}$ Constant, where the potential difference between the copper layers of the GEM foil was kept constant. The red line represents Case II: Electric field ($E$) Constant, where the potential difference between the copper layers was adjusted according to the copper thickness to maintain a constant electric field in every region. (a) illustrates the change in gain relative to the thickness of the lower copper layer in the GEM, while (b) depicts the fluctuations in ion backflow with the GEM lower copper thickness.}
    \label{fig:combined-figure}
\end{figure*}

Figure 12  shows how this ratio changes with the variation of the lower copper thickness of the foil. The X-axis of the plot represents the lower copper thickness and the Y-axis represents the ion backflow to gain ratio. The black line represents the scenario where the electric field changed with copper thickness, and the red line represents the scenario where the electric field was fixed despite copper thickness changes. It manifests that the ratio decreases as copper thickness increases from 2 \(\mu\)m to 10 \(\mu\)m, after which it begins to rise for Case I, where the voltage between the two copper layers is fixed. In the second scenario, where the electric field in each region is held constant, the ratio similarly decreases with increasing the copper thickness. For the existing GEM hole, the ratio between ion backflow and gain was 2.673. For single cone with lower copper thickness 5 \(\mu\)m, 10 \(\mu\)m, 15 \(\mu\)m, and 20 \(\mu\)m, the ratio of ion backflow to gain was 1.853, 1.437, 2.011 and 3.261 respectively, for case I. And for case II, the ratio was 1.853, 1.609, 1.383, and 1.583 respectively. 

\vspace{1 cm}

\textit{Table 1} illustrates how the ratio varies, either increasing or decreasing, with changes in copper thickness while keeping the existing GEM hole constant. A positive value indicates improved detector efficiency.

\begin{figure*}
  \centering
   {{\includegraphics[width=8.2cm]{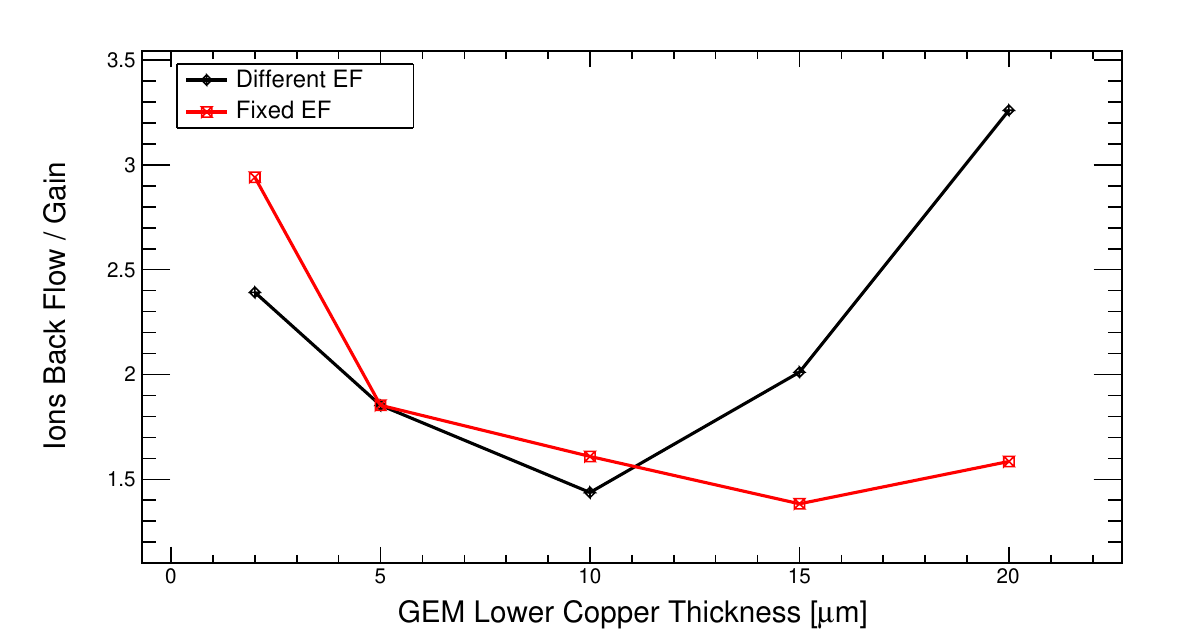} }}
    \caption{(color online) In studies using Garfield$^{++}$ to vary the lower copper thickness, the figure presents two cases. The black line, representing Case I, shows the scenario where the potential difference ($V_{GEM}$) across the copper layers of the GEM foil is kept constant. The red line, representing Case II, illustrates the situation where the potential difference is adjusted based on the copper thickness to ensure a constant electric field ($E$) throughout the regions. The figure displays how the ratio of ion backflow to gain fluctuates with changes in the GEM lower copper thickness.}
    \label{fig:1}
\end{figure*}

\FloatBarrier
\begin{table}
\centering
\begin{tabular}{|c|c|c|}
  \hline
  \multirow{2}{*}{Lower copper thickness in \(\mu\)m} & \multicolumn{2}{|c|}{ratio in \%} \\ \cline{2-3}
                            & Case I  & Case II \\ \hline
  05                    & 31\%      & 31\%      \\ \hline
  10                    & 46\%      & 40\%      \\ \hline
  15                    & 25\%      & 48\%      \\ \hline
  20                    & -22\%      & 41\%      \\ \hline
\end{tabular}
\caption{The increase of ions backflow and gain ratio for different lower copper thickness of single conical holes for the existing GEM hole in \%. A positive sign indicates the improvement and a negative sign indicates the deterioration of the efficiency of the detector.}
\label{tab:my_label}
\end{table}

\FloatBarrier
\section{Summary and Conclusion}
In this study, we have modeled the Gas Electron Multiplier and conducted a comprehensive investigation into the detector’s performance under various geometries and configurations. Key parameters such as pitch size, inner and outer hole diameters, and hole shape were systematically varied to assess their impact on the electric field distribution and detector performance. By analyzing these created configurations, we calculated the corresponding electric fields, providing insights into optimal geometries for efficient electron signal amplification. In addition to the geometric modeling using ANSYS, detailed simulations of electron transport and amplification processes were performed using Garfield$^{++}$. This combined approach provides a thorough understanding of how different geometries influence GEM detector performance.

For 10 events and single-line spectra, we have calculated the gain for both single-conical and bi-conical configurations for the GEM foil. The gain for the bi-conical hole is 15.0, while the single conical hole yielded a gain of 23.0, reflecting a 53\% increase. An extra-thick copper layer was introduced to reduce ion backflow and enhance structural durability. In Case II, where the electric field was uniform across all regions, the gain increased more rapidly with copper thickness compared to Case I, where $V_{GEM}$ was held constant. However, ion backflow also increased slowly with copper thickness. For Case I, the optimal ratio (ion backflow to the gain) was observed at 46\% with a 10 \(\mu\)m copper thickness for bi-conical holes. Case II, however, produced better gains and lower ratios, with ratios of 41\% and 42\% for copper thicknesses of 10 \(\mu\)m and 20 \(\mu\)m, respectively. Notably, a 10 \(\mu\)m copper layer is easier to etch than thicker layers to produce the GEM foil.

These findings suggest that the modified GEM geometry can achieve superior gain compared to standard designs. Increasing copper thickness not only reduces the ion backflow but also improves the durability of the GEM foil, enhancing both the performance and longevity of the detector. The study demonstrates that optimized GEM foil configurations can significantly boost the efficiency and reliability of the detector, making them more suitable for extended operational use.

\section*{Acknowledgments}
\addcontentsline{toc}{section}{Acknowledgment}
The authors gratefully acknowledge the Department of Science and Technology (DST, India) for funding this research through project EEQ/2020/000607, which made this work possible. A portion of this study was also presented at the HOT QCD Conference in 2024.

\FloatBarrier
\bibliographystyle{unsrt}
\bibliography{citation}
\end{document}